\documentstyle[aps,psfig,
twocolumn
]{revtex}

\begin{document}
\twocolumn[\hsize\textwidth\columnwidth\hsize\csname @twocolumnfalse\endcsname

\title{Phase diagram of the Kondo necklace: a mean-field
  renormalization group approach}

\author{Tatiana G. Rappoport and M. A. Continentino}
  
\address{Instituto de F\'\i sica, Universidade Federal Fluminense, \\
  CEP 24210-340, Niteroi RJ, Brazil; e-mail: tatiana@if.uff.br}
\date{Last modified: \today}

\maketitle

\begin{abstract}
  
  In this paper we investigate the magnetic properties of heavy
  fermions in the antiferromagnetic and dense Kondo phases in the
  framework of the Kondo necklace model. We use a mean field
  renormalization group approach to obtain a temperature versus Kondo
  coupling $(T-J)$ phase diagram for this model in qualitative
  agreement with Doniach's diagram, proposed on physical grounds. We
  further analyze the magnetically disordered phase using a two-sites
  approach. We calculate the correlation functions and the magnetic
  susceptibility that allow to identify the crossover between the
  spin-liquid and the local moment regimes, which occurs at a {\em
    coherence} temperature.
\end{abstract}

\vskip2pc]

\section{Introduction}
It is well known that the nature of the ground state of dense Kondo
compounds results basically from the competition between the
Ruderman-Kittel-Kasuya-Yosida (RKKY) interaction and the Kondo effect.
In a simple picture it is governed by a single parameter, the ratio
$J/W$, where $J$ is the effective exchange between localized moments
and conduction electrons and $W$ is the bandwidth of the latter.  The
value of this ratio is usually tunable experimentally by pressure or
composition ratio of the compounds.  The RKKY interaction is an
indirect magnetic interaction between localized moments, mediated by
the polarized conduction electrons, with an energy scale of order
$J_{RKKY}\propto \frac{J^2}{W}$, that produces a long-range ordered
magnetic ground state. On the other hand, the Kondo effect favors the
formation of singlet states between localized moments and conduction
electrons generating a non-magnetic ground state and, in the single
impurity case, has a characteristic energy scale of order
$k_BT_K=We^{-W/J}$. As result of the interplay between these two
effects, some Kondo compounds are non-magnetic and are characterized
by a heavy-fermion behavior (Fermi-liquid) at very low temperatures,
while others order magnetically, generally antiferromagnetically. The
study of this interplay is easily formulated using the Kondo lattice
model, which emphasizes the importance of spin fluctuations neglecting
charge fluctuations of the localized electrons, and has been well
characterized by the ``Doniach phase diagram''~\cite{doniach77}. In
this simple picture, the ordering temperature $T_N$ initially
increases with increasing $J$, then passes through a maximum and
vanishes at a critical coupling $J_c$. At this quantum critical point,
a second order phase transition between an antiferromagnetic ground
state for small values of $J$ and a dense Kondo state for strong
couplings $J$ occurs.  This behavior of $T_N$ has been experimentally
observed in many Cerium compounds by varying the pressure applied on
the system~\cite{croft79,eiling81,graf97} or the relative
concentration in the compounds~\cite{lee87,gignoux84}.

In this paper we are interested in studying the $T x J$ phase diagram
of Kondo compounds. For that purpose, we use an analog of the
symmetric Kondo lattice with complete absence of charge fluctuations,
the Kondo necklace model (KNM). This model was proposed by
Doniach~\cite{doniach77}, and its ground state has been investigated
by a variety of
methods~\cite{jullien77,scalettar85,santini92,moukouri95,zhang00}. In
order to assess the critical behavior of the Kondo necklace, we apply
the mean field renormalization group (MFRG), first proposed by Indekeu
{\it et al.}~\cite{indekeu82}, on the KNM. This method combines mean
field results for small clusters of spins and renormalization group
ideas. While mean field theory identifies the order parameter of the
cluster with the order parameter of the entire system, the MFRG
assumes that the cluster order parameter rescales with cluster
size~\cite{plascak99}.  Using this method we obtain the phase diagram
of the Kondo necklace as a function of temperature and Kondo coupling,
which is qualitatively identical to the Doniach diagram.  Within a
two-site approach~\cite{granato93}, we calculate the finite
temperature magnetic susceptibility and investigate the behavior of
short-range magnetic correlations in the magnetically disordered
phase.  As a result of this investigation, we increase the phase
diagram with a crossover line, which is associated to a coherence
temperature that separates Kondo spin-liquid and localized moments
regimes.
   
This paper is organized as follows: In the next section we present the
Kondo lattice and Kondo necklace models and discuss their analogies.
In section III we apply the mean-field renormalization group approach
in the Kondo necklace model and obtain the Doniach phase diagram,
which we compare with experimental results. In section IV we apply a
two-sites method on the non-magnetic phase, in order to study the role
of short-range correlations on the system. We also calculate the
magnetic susceptibility and then obtain a coherence temperature. In
section V we summarize and discuss our results.

\section{The model}

The Kondo lattice model (KLM) is a theoretical model for heavy
fermions that can be derived from the more fundamental Anderson
Lattice Model, in the case of well-developed local spin
moments~\cite{schrieffer66}. It consists of two different types of
electrons, the localized spins whose charge degrees of freedom are
suppressed, and the conduction electrons that propagate as charge
carriers. It is described by

\begin{equation}
H=-t\sum_{{\bf \langle }i,j{\bf \rangle }}(c_{i,\sigma }^{\dagger
}c_{j,\sigma }+h.c.)+J\sum_i{\bf S}_i\cdot c_{i,\alpha }^{\dagger }
{\boldmath{\sigma }}_{\alpha \beta }c_{i,\beta }. 
\label{Kondo}
\end{equation}

The first term represents the conduction band ($
c_{i,\sigma}^{\dagger}$ is the creation operator, $t$ is the hopping
between nearest neighbors) and the second term is the interaction
between conduction electrons and localized moments ${\bf S}_i$ via the
intra-site exchange $J$, where $\sigma$ are the Pauli matrices.

In order to study the interplay between Kondo screening and the RKKY
interaction, Doniach proposed a simplified model related to the
one-dimensional Kondo lattice, called the Kondo necklace model (KNM).
In this model, the conduction electrons are replaced by a spin chain
with $XY$ coupling which eliminates charge
fluctuations~\cite{doniach77}:
\begin{equation} 
H_{KN}=W\sum_{\langle i,j\rangle }(\tau _i^x\tau _j^x+\tau
_i^y\tau _j^y)+J\sum_i{\bf S}_i\cdot {\boldmath{\tau} }_i, 
\label{necklace}
\end{equation}
where ${\bf \tau }_i$ and ${\bf S}_i$ are independent sets of spin
$1/2$ Pauli operators.  The first term mimics electron propagation,
and in one dimension can be mapped by the Jordan-Wigner transformation
onto a band of spinless fermions. The second term is the magnetic
interaction between conduction electrons and localized spins ${\bf
  S}_i$ via the coupling $J$, as in Eq. (\ref{Kondo}).

Although the KLM is mapped onto the KNM only in one dimension, it is
clear that, even in higher dimensions, $H_{KN}$ has the same magnetic
tendencies of the Kondo lattice. Since the essential features of the
original model are kept, we expect that the main physical properties
of the Kondo lattice will be maintained in the model described by Eq.
(\ref{necklace}).

Since in the KNM approach to the Kondo lattice charge fluctuations are
neglected, the critical properties in this case are described just by
spin excitations. However, it is important to emphasize, that the
analysis of heavy fermions systems in terms of the Kondo necklace
model is appropriate~\cite{godart89}. A recent and very complete study
of a heavy fermion system just at the quantum critical point QCP
~\cite{schroder00} shows that a description in terms of local moments
seems to be more appropriate for this kind of material. This may not
be the case for all heavy fermions but it is also true that the
universality class of the quantum transition of heavy fermions has not
been determined yet and may even not be unique.

\section{Mean-field renormalization group}

The mean-field renormalization group was first proposed by Indekeu
{\it et al.}~\cite{indekeu82} for computing critical properties of
lattice spin systems. This method has been applied to many statistical
physics problems, both classical and quantum systems,
with~\cite{droz82} and without disorder~\cite{indekeu82}, and the
resulting critical exponents deviate from those obtained from standard
mean field theories (including Bethe lattice
calculations)~\cite{plascak99}.

The main idea of the MFRG is the comparison of two clusters with $N$
and $N'$ sites respectively, subjected to symmetry-breaking boundary
conditions. The interactions within the clusters are treated exactly,
and the effect of surrounding sites is replaced by a mean field which
is supposed to scale in the same way as the order parameter.

For a generic spin system, one considers each boundary spin fixed and
equal to $b$ and $b'$ for the $N$ and $N'$ spin clusters,
respectively.  After computing the order parameter ${\cal O}_N$ and
${\cal O}_{N'}$ for both clusters one imposes a scaling relation
between them
\begin{equation}
{\cal O}_{N'}(K',b')=\xi {\cal O}_N(K,b),
\label{scaling}
\end{equation}
with $K$ and $K'$ being the coupling constants of the two rescaled
systems. Assuming a similar relation between the mean-fields ($b'=\xi
b$) and knowing that these fields have to be very small near the
second order phase transition, one can expand equation (\ref{scaling})
for small values of $b$ and $b'$ to obtain
\begin{equation} 
\left .\frac{\partial{{\cal O}_N'}(K',b')}{\partial b'}\right|_{b'=0} = 
\left .\frac{\partial{{\cal O}_{N}}(K,b)}{\partial b}\right|_{b=0}~,
\label{rect}
\end{equation}  
which is independent of the scaling factor $\xi$. Equation
(\ref{rect}) can be interpreted as a recursion relation for the
coupling constant $K$, from which the critical point $K_c$ is
extracted.

\begin{figure}[!htb] 
\centerline{
\psfig{figure=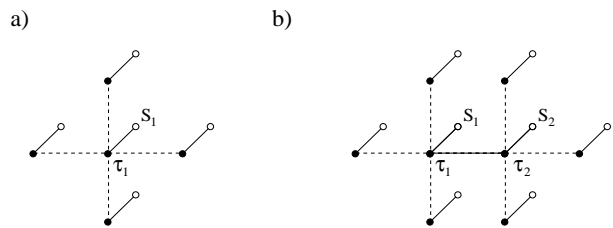,width=8cm}}
\vskip 0.5cm
\caption{ Clusters considered in the calculations of the $2D$ case. 
 Each site contains two spins, one spin $S$ (open circle)
  connected by a solid line to one spin $\tau$ (filled circle). The
  interaction between sites is mediated by the spins $\tau$, where the
  dashed bonds represent interactions with the boundary symmetry
  breaking fields $b'$ and $b$.}
\label{fig1}
\end{figure}

We can apply this method to the Kondo necklace in its simplest
version, that is, we consider two cells containing one and two sites
each, as sketched in figure \ref{fig1} for a $2D$ hipercubic lattice.

As we are dealing with an incipient antiferromagnetic ordering and the
antiferromagnetism in the Kondo necklace occurs in the $XY$
plane~\cite{doniach77}, we consider a D-dimensional hipercubic lattice
and divide the system in two sub-lattices $A$ and $B$. The order
parameter is then the staggered magnetization of the spins $\tau$,
taken along the $x$ direction (see Eq. (\ref{H1}) and (\ref{H12})
below). The $x$ component of the boundary spins in the smallest
cluster is fixed to be $-b'$ since all the first neighbors of $\tau_1$
are in the same sub-lattice. In the two-sites cluster, $\tau_1$ and
$\tau_2$ are in different sub-lattices, so the $x$ component of their
neighboring boundary spins have different signs and are fixed at $-b$
and $b$ respectively.

Let us first consider the Hamiltonian for a one-site cluster taken on
a sub-lattice A:

\begin{equation} 
H_{1}=J'~{{\bf S}_{1}}\cdot{{\bf \tau}_{1}}-zb'W'\tau^{x}_{1},
\label{H1} 
\end{equation} 
\noindent where $J'$ is the scaled coupling interaction, and the spin $\tau_1$
interacts with its $z$ nearest neighbors through the term
$zb'W'\tau^{x}_{1}$.  

Similarly, the Hamiltonian for the two-sites
system (one in each sub-lattice) is given by:
\begin{equation}
H_{12}=J\sum_{i=1}^{2}{{{\bf S}_{i}}\cdot{{\bf \tau}_{i}}}+W(\tau
_1^x\tau _2^x)-(z-1)bW(\tau^{x}_1-\tau^{x}_2).
\label{H12}
\end{equation}

In this case the spin $\tau_1$ interacts directly with $\tau_2$
through a term $W(\tau_1^x\tau _2^x)$ and both $\tau_1$ and $\tau_2$
interact with their $(z-1)$ nearest neighbors through
$-(z-1)bW\tau^{x}_1$ and $(z-1)bW\tau^{x}_2$, respectively.

This method can be used to study quantum systems ($T=0$), for which
there are few available renormalization group techniques. For that
purpose, we compute the ground state $|0>_{1}$ and $|0>_{12}$ of the
two systems and their corresponding staggered magnetizations along the
$x$ direction. In the vicinity of the phase transition, $b$ and $b'$
can be assumed small and
\begin{eqnarray}
{M^{s}_{1}}_&=&<0|\tau^{x}_{1}|0>=-\frac{z}{2j'}b',
~~~~~~~~~~~~~~~~~~~~~~~~~~~~~~~~~~ ~~ \nonumber
\end{eqnarray}
\begin{eqnarray}
{M^{s}_{12}}&=&\frac{<0|(\tau^{x}_{1}-\tau^{x}_{2})|0>}{2}\nonumber \\
&=& -\frac{2\,(z - 1)\,( \sqrt{16\,j^{2} + 1} +
1)^{2}}{(1 + 16\,j^{2} +\sqrt{ 16\,j^{2} + 1})\,( - 1 +
\sqrt{16\,j^{2} + 1})}b,\nonumber
\end{eqnarray}   
where $j=\frac{J}{W}$ and $j'=\frac{J'}{W'}$.

The main assumption of the MFRG is the imposition of the same scaling
relations between $M^{s}_{1}$ and $M^{s}_{12}$, $b$ and $b'$. By doing
this we arrive at the renormalization group recursion relation for $j$
and $j'$. The associated fixed point equation is
\begin{equation}
\frac{z}{j_{c}}=\frac {4(z - 1)( \sqrt{16\,j_c^{2} +
 1} + 1)^{2}}{(1 + 16\,j_c^{2} +\sqrt{ 16\,j_c^{2} + 1})\,( -
 1 + \sqrt{16\,j_c^{2} + 1})}.
\label{rect0}
\end{equation}

By solving Eq. (\ref{rect0}) we obtain the antiferromagnetic quantum
critical point (QCP) in any space dimension $D$. For $z=2$, which
corresponds to a one dimensional system, we obtained $j_c=0.64$, The
critical ratios $j_c$ for other coordination numbers are:
$z=4\rightarrow j_c=1.66$, $z=6\rightarrow j_c=2.67$. This values can be
compared with recent results~\cite{assad99,zhang00} and the relative
errors are $12.67\%$ for 2D and $1.5 \% $ for 3D showing that the results are
most relievable for higher dimensions.

It is worth mentioning that the diagonalization of $H_{12}$ is not a
simple task. However, as we are interested in small values of the mean
field, a perturbative expansion can be worked out in order to obtain
the eigenvalues and eigenvectors in powers of $b$~\cite{plascak84}. We
follow the same prescription in order to calculate the staggered
magnetizations for $T\neq 0$.

As we have more then one variable, we cannot determine a complete
renormalization flow in the $[j,T]$ plane. However, for a fixed $j$ we
can calculate $T_N$ or vice-versa. In order to obtain the critical
values, we consider the solutions of the fixed point equation
associated with eq. (\ref{rect}), that is,
\begin{equation}
\left .\frac{\partial{M^{s}_{1}}(T_N,j)}{\partial
b'}\right|_{b'=0} =
\left .\frac{\partial{M^{s}_{12}}(T_N,j)}{\partial
b}\right|_{b= 0}~,
\end{equation} 
where for each $j<j_c$ we obtain a $T_N(j)$ and the set of all
critical points yields the temperature dependent phase diagram of the
Kondo necklace. The critical line for a $3D$ system ($z=6$),
is depicted in figure \ref{fig1}. Close to the zero temperature fixed
point it behaves as:
\begin{equation}
|\delta|=|j-j_c| \propto e^{-\alpha(z,j_c)\beta_N}, ~~~~~\beta_N=\frac{1}{k_BT_N}
\end{equation}
This dependence is characteristic of this
approach and appears for any dimension.

\begin{figure}[!htb] 
{\centerline {\psfig{figure=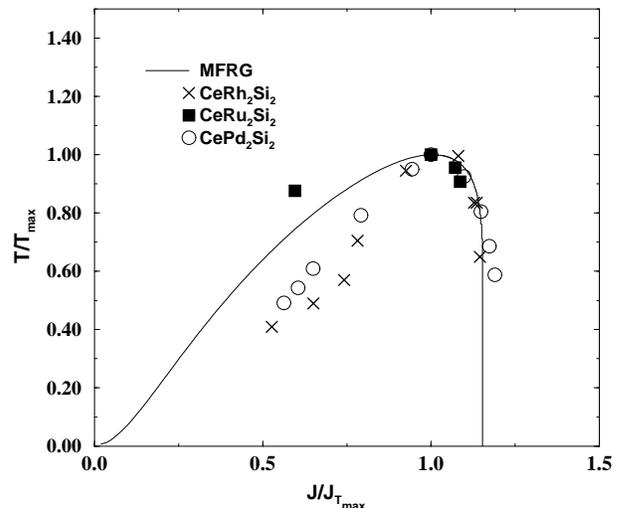,width=8cm,angle=270}}
\caption{ Phase diagram of the KNM within the mean field renormalization group
approach (with $z=6$) compared with experimental results for Cerium compounds
  extracted from ref.\protect\onlinecite{cornelius94}. The results are
  normalized by the maximum value of the Ne\`el Temperature $T_N$ and its
  associated coupling $J$.}}
\label{fig2}
\end{figure}

In figure \ref{fig2}. Endstra {\it et al.} calculated the $J$ coupling of
many Cerium and Uranium compounds based fundamentally on the atomic radii and
interatomic distances of these systems~\cite{endstra93}. More recently,
Cornelius and Schilling, based on the variation of the lattice parameters
generated by pressure and substitution of $Si$ by $Ge$ (negative pressure) in
$CeM_{2}Si_{2-x}Ge_{x}$ compounds (where M=Rh,Ru,Pd)~\cite{rh,ru,pd}, found a
relation between the Ne\`el temperature $T_N$ and the coupling $J$ for
$CeRu_2Si_2$, $CeRh_2Si_2$ and $CePd_2Si_2$~\cite{cornelius94}. Their results
fall in Doniach-like curves, that when normalized by the maximum of $T_N$
($T_{max}$) and its equivalent $J_{T_{max}}$ collapse onto a single universal
curve. 

We observe that in the experimental results, the Ne\`el temperature
goes to zero in the weak coupling regime at a value of $J$ greater than zero,
differently from the theoretical predictions for the Kondo Lattice.  By
shifting the experimental results along the $J$ axis in order to obtain
$T_N=0$ at $J=0$, and using the same normalization used in
Ref.\onlinecite{cornelius94}, we obtain the results of figure \ref{fig1} ($z=6$ ($3D$)), 
where
one can see that the behavior of the theoretical curve is in good agreement
with the experimental values.  We can conclude that these Kondo systems behave
qualitatively as proposed by Doniach although in practice the long range
magnetic order vanishes before the theoretical $J=0$ value is attained.

As we are dealing with a mean-field like approach using small clusters
we do not  obtain a precise value for the QCP for one
dimension case ($J=0$) such as in previous
works~\cite{scalettar85,santini92,moukouri95,zhang00}.  Nevertheless,
this method has the advantage of being very simple to implement at
finite temperature and produce relievable values of the QCP for higher
dimensions, as discussed before.

The main feature of our MFRG method in its simplest approach ( 1 and 2
site cells) is that it yields the $[T,j]$ phase diagram for heavy
fermions as proposed by Doniach on physical grounds~\cite{doniach77}
which, as far as we know, has never been obtained before. This is so
because this method captures the essential physics of the Kondo
lattice problem, namely, the competition between the RKKY and Kondo
interactions. However, it is important to note that the present method
can be applied to larger clusters, which provides better information
concerning the lattice topology of the system~\cite{plascak99}, in
order to obtain more accurated values for the quantum critical point for
lower dimensions.

\section{Short-range correlations}  

\begin{figure}[!htb]
a) 
\centerline{\psfig{figure=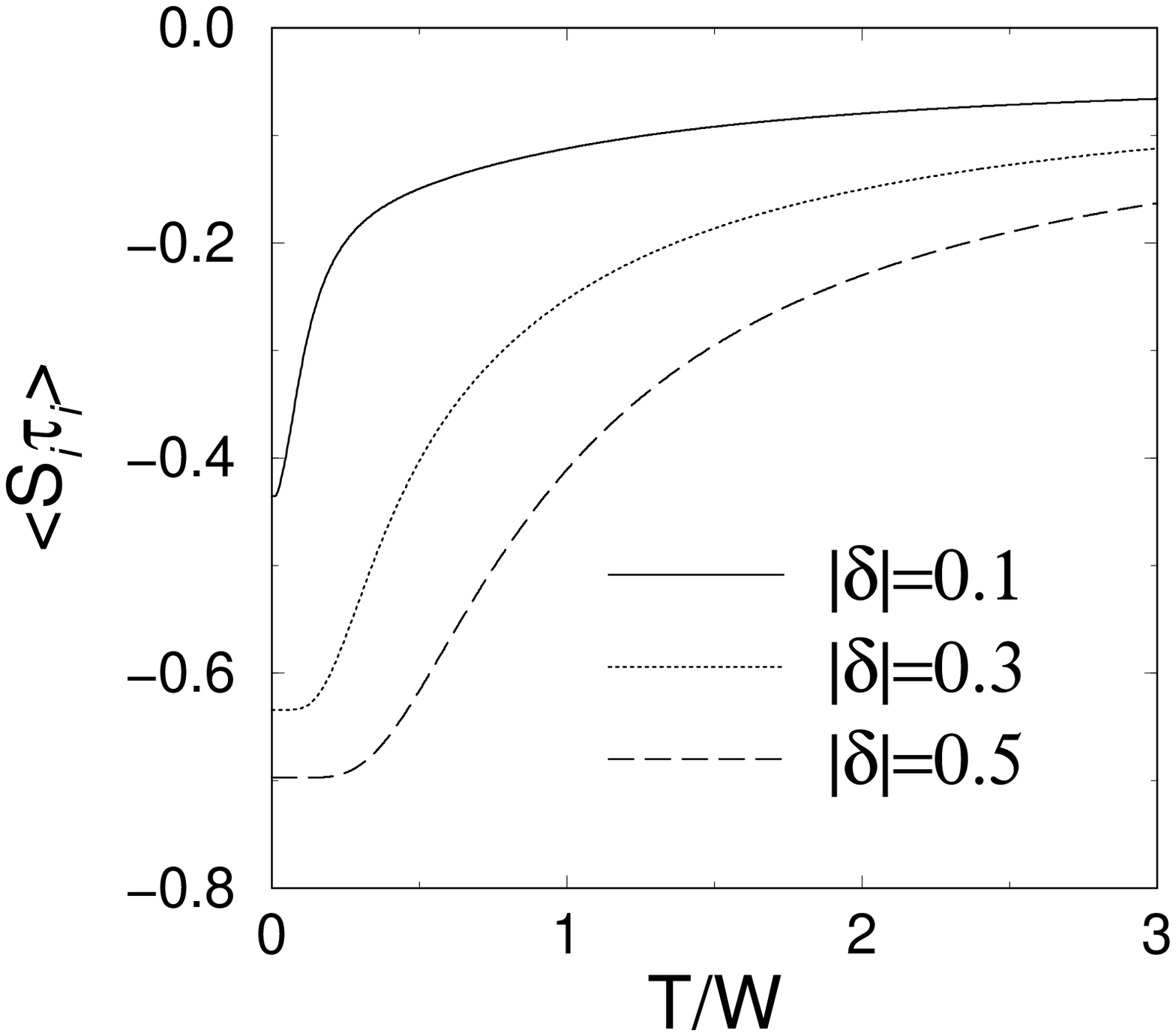,width=8cm,height=4cm}}
b)
\centerline{\psfig{figure=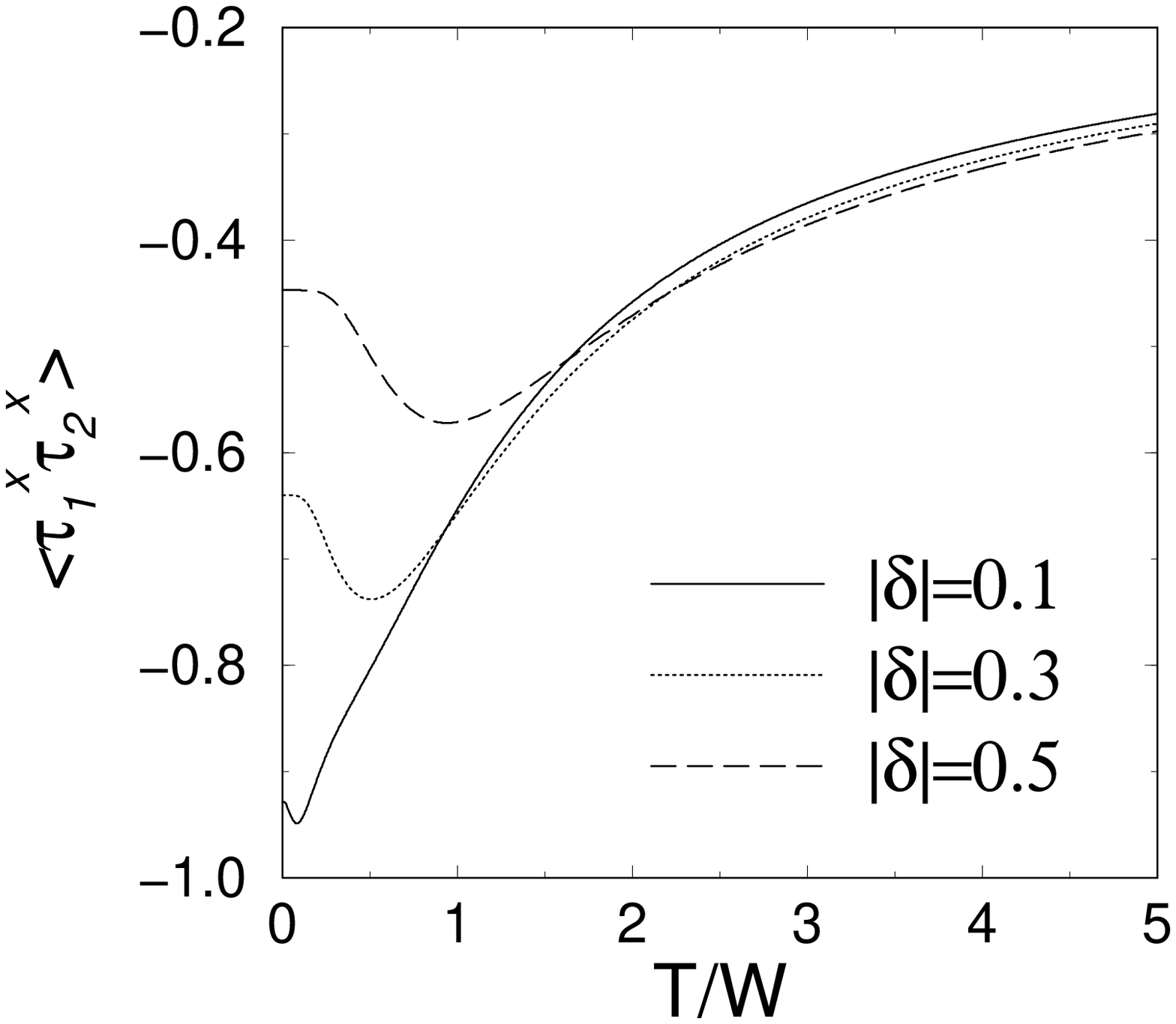,width=8cm,height=4cm}}
\vskip 0.5cm
\caption{{} Correlation functions $<S\cdot\tau>$ (a) and
  $<\tau^{x}_1\tau^{x}_2>$ (b) for
  different values of $|\delta|=|j-j_c|$ as a function of temperature.}
\label{fig3}
\end{figure}

In the magnetically disordered phase, where the mean-fields $b$ and $b'$ are
identically zero, we can still calculate some properties of the system by
means of a two-site approximation. This method consists on solving exactly a
Hamiltonian of two sites, i.e., Eq. (\ref{H12}) with $b=0$. As the system does
not present long-range (magnetic) order, we expect that this approach will 
 unravel some aspects of this phase. We can investigate the behavior of
short-range magnetic correlations as well as the intra-site correlations
between localized moments and conduction electrons, which are related to the
Kondo effect. The results of the calculations are shown in figures
\ref{fig3}(a) and \ref{fig3}(b).

The localized-delocalized spin correlation function $<S\cdot\tau>$
characterizes the condensation of singlet states. At $T=0$, this function
decreases for increasing values of $j$, saturating at the value
$-\frac{3}{4}$, which is consistent with the strong coupling
limit~\cite{moukouri95}, where the system condenses into independent singlets
in each site. As the temperature increases, this correlation rapidly falls to
zero and we observe that, since coherence in the KNM is a collective
phenomena, even low temperature excitations can destroy the Kondo spin-liquid
regime.  The inter-site correlation function $<\tau^{x}_1\tau^{x}_2>$
represents the short-range magnetic correlations. This function has a minimum
at a low temperature and then gradually vanishes as T
increases. Differently from $<S\cdot\tau>$, it has a smooth behavior for large
$T$ and its asymptotic behavior is independent of $|\delta|=|j-j_c|$.

The competition between these two kind of correlations produces a
change in the behavior of the system, as illustrated in figure
\ref{fig3}. For low temperatures, the correlation $<S\cdot\tau>$ is
stronger and the system is in a condensate of singlets, typical of the
Kondo spin-liquid regime~\cite{zhang00}. Near the minimum of the
magnetic correlation function $<\tau^{x}_1\tau^{x}_2>$ there is a
crossover: $<\tau^{x}_1\tau^{x}_2>$ begins to dominate and destroys
these singlet states giving rise to a paramagnetic regime with
localized moments.  The inflection point in the magnetic correlation
function, where it crosses $<S\cdot\tau>$, defines the crossover or
coherence temperature which separates two different regimes in the
non-magnetic region of the phase diagram.

\begin{figure}[!htb] 
\centerline{
\psfig{figure=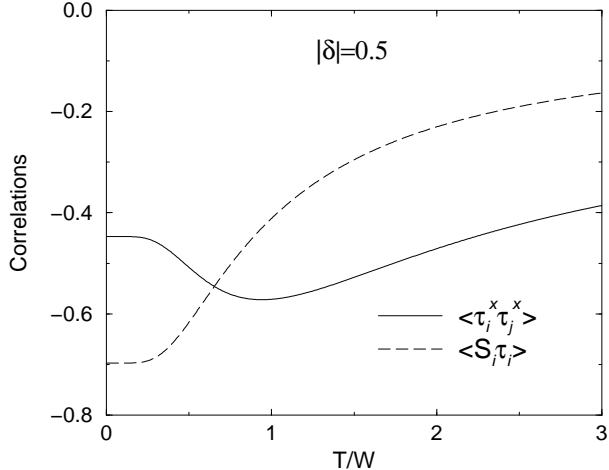,width=8cm}
}
\vskip 0.5cm
\caption{{} Competition between two kind of
  correlations: the intra-site $<S\cdot\tau>$ (dashed line) and the
  inter-site $<\tau^{x}_1\tau^{x}_2>$ (solid line) for
  $|\delta|=0.5$.}
\label{fig4}
\end{figure}

The magnetic susceptibility (figure \ref{fig4}), calculated according to
eq.(\ref{chi}), illustrates the difference between these two regimes.
\begin{equation}
\chi_0=Z^{-1}N^{-1}\beta\sum_m{e^{-\beta
E_m}(M^{x}_{total})^2}, 
\label {chi} 
\end{equation}
$$
Z=\sum_m{e^{-\beta E_m}}, ~~\beta=\frac{1}{k_BT}.
$$
At low temperatures, as we are dealing with the symmetric case of
the Kondo lattice, the system is an insulator and there is a
singlet-triplet gap. At a higher temperature there is a maximum in
the susceptibility where the gap closes and finally, at high $T$, we
can see an asymptotic Curie-Weiss regime, typical of localized
moments. For $|\delta|=0$ the system has only the Curie-Weiss regime
and the susceptibility can be fitted as, $\chi_0 \propto T^{-1}$,

The coherence temperature can be calculated by finding the zero of
$\frac{\partial^2<\tau^{x}_1\tau^{x}_2>}{\partial T^2}$ and at low
temperatures it is quadratic in $|\delta|$:
\begin{equation}
T_{coh}\propto |\delta|^{2}.
\end{equation}
The gap at $T=0$ in the critical regime, for $|\delta|\rightarrow 0$,
also has a power law behavior with the same exponent of the coherence
temperature.

\begin{figure}[!htb] 
\centerline{
\psfig{figure=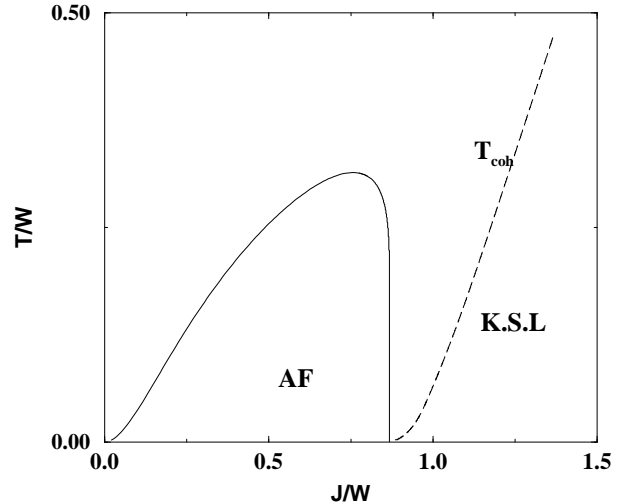,width=8cm}
}
\vskip 0.5cm
\caption{{} Complete phase diagram of the Kondo necklace model (for
$z=6$). The curve $T_N(J)$ is shown here as a full line, whereas the dashed
line represents $T_{coh}(J)$.}
\label{fig5}
\end{figure}

The Doniach diagram discussed in the previous section can be
generalized in order to include two other regimes in the magnetically
disordered region of the phase diagram, separated by the coherence
line. Figure \ref{fig5} illustrates the extended diagram obtained
using these two methods.

\section{Conclusions} 

We have successfully applied the MFRG on the Kondo necklace model.
Using the simplest choice of clusters, we obtained the quantum
critical point for the $2D$ and $3D$ cases that are in good agreenment
with previous calculations. We obtained for the first time a
theoretical calculation of the Doniach diagram and compared it with
experimental results for some Cerium compounds obtaining qualitatively
good agreement. In the magnetic disordered phase we calculate, using
a two-sites method, the short-range magnetic correlation function, the
localized-delocalized spin correlation function as well as the
magnetic susceptibility as a function of temperature and characterized
two well-defined regimes: at low temperatures, a condensate of
singlets with a singlet-triplet gap and at high temperatures, a
Curie-Weiss regime.  The crossover line close to the QCP has been
analytically calculated and presents a power-law behavior.

\section{Acknowledgments}
The authors wish to thank the Brazilian agencies CNPq and FAPERJ for
financial support and Dr. E. Miranda for useful discussions at the
initial stage of this work.

\end{document}